\documentclass{aa}
\usepackage{graphicx}
\usepackage{amssymb}
\begin{document}
   \title{Stellar Evolution with Rotation and Magnetic Fields}

   \subtitle{IV: The Solar Rotation Profile} 

   \author{P. Eggenberger \and A. Maeder \and G. Meynet}
   \institute{Observatoire de Gen\`eve, 51 chemin de Maillettes,
             CH-1290 Sauverny, Switzerland}
       
   \offprints{Patrick Eggenberger}
   \mail{Patrick.Eggenberger@obs.unige.ch}
  
  \date{Received ; accepted }
  \abstract{We examine the generation of a magnetic field in a solar-like star and its effects on the internal
  distribution of the angular velocity. We suggest that the evolution of a rotating star with magnetic fields
  leads to an equilibrium value of the differential rotation. This equilibrium is determined by the magnetic
  coupling, which favours a constant rotation profile, and meridional circulation which tends to build
  differential rotation. The global equilibrium stage is close to solid body rotation
  between about 0.7 and 0.2\,$R_{\odot}$, in good agreement with helioseismic measurements. 
}
  \maketitle
%
%________________________________________________________________

\section{Introduction}

One of the most severe problems in stellar physics concerns the rotation profile of the radiative interior of the Sun. 
Helioseismological results indicate that the angular velocity $\Omega(r)$ is constant as a function of the radius $r$
between about 20\,\% and 70\,\% of the total solar radius $R_{\odot}$ (Brown et al. \cite{br89}; Kosovichev et al. \cite{ko97};
Couvidat et al. \cite{co03}), while meridional
and rotational turbulent diffusion (e.g. Zahn \cite{za92}) produce an insufficient internal coupling to ensure
solid body rotation (Pinsonneault et al. \cite{pi89}; Chaboyer et al. \cite{ch95}). This suggests that another effect intervenes. Mestel \& Weiss (\cite{me87})
proposed that a weak internal magnetic field could provide the required internal coupling. 
Charbonneau \& MacGregor (\cite{ch93}) also showed that large-scale internal magnetic fields can yield a weak
internal differential rotation at the solar age. 
Another proposition to explain
the flat rotation profile of the Sun is the angular momentum transport by internal gravity waves 
(Zahn et al. \cite{za97}; Talon et al. \cite{ta02}; Talon \& Charbonnel \cite{ta05}). 
 
An efficient dynamo has been proposed to operate in stellar radiative layers in differential rotation
(Spruit \cite{sp02}). This dynamo is based on the Tayler instability, which is the first one to occur in a radiative zone 
(Tayler \cite{Tay73}; Pitts \& Tayler \cite{Pitts86}). Even a very weak horizontal 
magnetic field is subject to Tayler instability, which then creates a vertical field component,
which is wound up by differential rotation. As a result, the field lines become progressively
closer and denser and thus a strong horizontal field is created at the
energy expense of differential rotation.
 
Maeder \& Meynet (\cite{Magn1}) studied the effects of the Spruit dynamo on the evolution of massive stars. 
They showed that a magnetic field can be created during the main sequence evolution of a rotating star 
by the Spruit dynamo and examined the timescale for the field 
creation, its amplitude and the related diffusion coefficients. Maeder
\& Meynet (\cite{Magn2}) also developed a generalisation of the
dynamo equations in order to encompass all cases of $\mu$-- and $T$--gradients, 
as well as all cases from the fully adiabatic to non--adiabatic solutions. 
The clear result of these studies is that magnetic
field and its effects are quite important in massive stars. The magnetic coupling, resulting from the
Tayler--Spruit dynamo, is found to be able to enforce solid body rotation in massive stars. The question 
is now whether this dynamo also operates efficiently in a slow rotator like the Sun, so as to be responsible
for the observed constancy of $\Omega(r)$.  
  
In Sect. 2, we collect in a short consistent way the basic equations of the dynamo.
In Sect. 3, the numerical models are presented, while Sect. 4 gives the conclusion.

%-----------------------------------------------------------------------------
\section{The dynamo equations}  
\label{dyn}

In this section we briefly summarize the consistent system of equations for the dynamo 
(see Spruit \cite{sp02} and Maeder \& Meynet \cite{Magn2} for more details).

The energy density $u_{\mathrm{B}}$ of a magnetic field of intensity $B$ per volume unity is
  \begin{eqnarray}
u_{\mathrm{B}} = \frac{B^2}{8  \; \pi} \,  = \,{1 \over 2} \; \rho \; r^2 \omega_{\mathrm{A}}^2 \,
\quad \mathrm{with} \quad	\omega_{\mathrm{A}}= \frac{B}{(4\; \pi \rho)^{1\over 2} \;r}  \;	,
\end{eqnarray}%Eq.1

  \noindent
where $\omega_{\mathrm{A}}$ is  the  Alfv\'en frequency   in a spherical geometry.
In stable radiative layers, there is in principle no particular motions. However, if due to
magnetic field or rotation, some  unstable displacements
 of vertical amplitude $l/2$ occur around an average stable position,
 the restoring buoyancy force produces  vertical oscillations around the equilibrium position with a
frequency equal to the Brunt--V\H{a}is\H{a}l\H{a} frequency $N$ (see Eq.~14 of Maeder \& Meynet \cite{Magn2}
for the definition of $N$ in a medium with both thermal and magnetic diffusivity $K$ and $\eta$).

The restoring oscillations will have an average density of kinetic energy 

\begin{equation}
u_{\mathrm{N}}  \simeq  \; f_{\mathrm{N}} \; \rho \; l^2 \; N^2  \; ,
\label{uN}
\end{equation}%Eq.4

\noindent
where $f_{\mathrm{N}}$ is  a geometrical factor of the order of unity.
If the magnetic field produces some instability with a vertical component, one must have
 $u_{\mathrm{B}} >  u_{\mathrm{N}}$.
 Otherwise, the restoring force of gravity which acts at the dynamical timescale
 would immediately counteract the magnetic instability. From this inequality, one obtains 
$ l^2 < \frac{1}{2f_{\mathrm{N}}} \;r^2 \; \frac{\omega^2_{\mathrm{A}}}{N^2}$.
 If $f_{\mathrm{N}}= {1 \over 2}$, we  have the condition for the vertical amplitude of the  instability
 (Spruit \cite{sp02}; Eq.~6),
 
 \begin{equation}
 l <  \; r \; \frac{\omega_{\mathrm{A}}}{N}  \; ,
 \label{lr}
 \end{equation}%Eq.5
  \noindent
  where  $r$ is the radius.
  This means that there is a maximum size of the vertical length $l$ of a magnetic instability. In order to not be quickly  damped 
  by magnetic diffusivity, the vertical length scale of the instability must satisfy 
  
  \begin{equation}
 l^2 >  \frac{\eta}{\sigma_{\mathrm{B}}}= 
 \frac{\eta \; \Omega} {\omega_{\mathrm{A}}^2}  \; , 
 \label{lmin}
 \end{equation} % Equ.6
 
 \noindent
 where $\Omega$ is the angular velocity and  
 $\sigma_{\mathrm{B}}$ the characteristic growth-rate of the magnetic field.
 In a rotating star, this growth-rate is  $\sigma_{\mathrm{B}} =
 (\omega_{\mathrm{A}}^2 / \Omega)$  due to the  Coriolis force (Spruit \cite{sp02}; see also Pitts \& Tayler \cite{Pitts86}). 
The combination of the limits given by Eqs. (\ref{lr}) and (\ref{lmin}) gives for the case of marginal stability,

\begin{equation}
\left(\frac{\omega_{\mathrm{A}}}{\Omega}\right)^4  =  \frac{N^2}{\Omega^2} \;
\frac{\eta}{r^2 \; \Omega}   \; .
\label{premier}
\end{equation} %Equ.7

The equality of the amplification time of Tayler instability
  $\tau_{\mathrm{a}}=  N /(\omega_{\mathrm{A}} \Omega q)$  with the  characteristic frequency 
  $\sigma_{\mathrm{B}}$ of the magnetic field leads to the equation (Spruit \cite{sp02})
  
\begin{eqnarray}
\frac{\omega_{\mathrm{A}}}{\Omega} = q \; \frac{\Omega}{N}
  \quad
\mathrm{with} \quad q= -\frac{\partial \ln \Omega}{\partial \ln r}  \; .
\label{qo}
\end{eqnarray} %Eq.8
\noindent

By  eliminating  the expression of $N^2$ between Eqs. (\ref{premier}) and (\ref{qo}), we obtain an expression for the magnetic diffusivity, 

\begin{equation}
\eta \;= \; \frac{r^2 \; \Omega}{q^2} \; \left( \frac{\omega_{\mathrm{A}}}
{\Omega}\right)^6  \; .
\label{eta}
\end{equation}%eq.10

\noindent
Eqs. (\ref{premier})  and (\ref{qo}) form a coupled system relating the two unknown 
quantities $\eta$ and $\omega_{\mathrm{A}}$. Instead, one may also consider for example the system
formed by Eqs. (\ref{qo}) and (\ref{eta}).
Formally, if one accounts for the complete expressions of the thermal gradient $\nabla$, the system of equations would be
of degree 10 in the unknown quantity  $x=\left(\frac{\omega_{\mathrm{A}}}{\Omega}\right)^2$ (Maeder \& Meynet \cite{Magn2}).  
The fact that the ratio $\eta/K$ is very small allows us to bring these coupled equations to a system of  degree 4 (Maeder \&
Meynet \cite{Magn2}),

\begin{equation}
\frac{r^2 \Omega}{q^2 K} \left(N_{\mathrm{T}}^2 + N_{\mu}^2 \right)  x^4-
\frac{r^2 \Omega^3}{K} x^3 + 2 N_{\mu}^2 \; x - 2 \Omega^2 q^2 = 0 \; .
\label{equx}
\end{equation} %Eq.11

\noindent
 $K$ is the radiative diffusivity, namely  $K= \frac{4ac T^3}{3 \kappa \rho^2 C_{\mathrm{p}}}$. The 
 solution of this equation, which is easily obtained numerically, provides the 
 Alfv\'en frequency and by Eq. (\ref{eta}) the thermal diffusivity.

The azimuthal component of the magnetic field is much stronger that the radial one in the Tayler--Spruit dynamo.
We have for these components (Spruit \cite{sp02})

 \begin{equation}
  B_{\varphi}= (4 \pi \rho)^{\frac{1}{2}} \; r \; \omega_{\mathrm{A}} \quad \mathrm{and} \quad
 B_ {\mathrm{r}}= B_{\varphi} \; (l_{\mathrm{r}} / r)  \; ,
 \label{champ}
 \end{equation}%Eq.13
 
 \noindent
 where $\omega_{\mathrm{A}}$ is the solution of the general
 equation (\ref{equx}) and $l_{\mathrm{r}}$ is given by Eq.~(\ref{lr}).

 Turning towards the transport of angular momentum by  magnetic field, we first write
  the azimuthal stress  by volume unity due to the magnetic field  
  
  \begin{eqnarray}
  S \; = \frac{1}{4 \; \pi} \; B_{\mathrm{r}} B_{\varphi} \; = \;
  \frac{1}{4 \; \pi} \;  \left(\frac{l_{\mathrm{r}}}
  {r}\right) B_{\varphi}^2 = % \nonumber \\[2mm]
  \; \rho \; r^2 \; \left(\frac{\omega_{\mathrm{A}}^3}{N}\right)  \; .
  \end{eqnarray}%Eq.14

 \noindent
 Then, the viscosity $\nu$
 for the vertical transport of angular momentum can be expressed in terms of  $S$
 (Spruit \cite{sp02}),
 
 \begin{equation}
 \nu = \frac{S}{\rho \; q \; \Omega} =
  \; \frac{\Omega \; r^2}{q} \;
 \left( \frac{\omega_{\mathrm{A}}}{\Omega}\right)^3 \; 
\left(\frac{\Omega}{N}\right) \; .
 \label{nu}
 \end{equation}%Eq. 15
 
 \noindent
 This is the general expression of $\nu$ with $\omega_{\mathrm{A}}$ given by the solution
 of Eq.~(\ref{equx}).   
 We have the full set of expressions necessary  to obtain  
 the Alfv\'en frequency $\omega_{\mathrm{A}}$ and  the magnetic diffusivity  $\eta$. Let us recall that
 $\eta$   also
 expresses  the vertical transport of the chemical elements, while the viscosity $\nu$
  determines the vertical transport of angular momentum by the magnetic field.

\section{Stellar models}

We consider here models of 1\,M$_{\odot}$, with the solar chemical composition of Grevesse \& Sauval (\cite{gr98}).
The stellar evolution code used for these computations is the Geneva code
including shellular rotation (Meynet \& Maeder \cite{MMXI}). 
We use the braking law
of Kawaler (\cite{ka88}) in order to reproduce the magnetic braking undergone by low
mass stars when arriving on the main sequence. Two parameters enter this braking law: the
saturation velocity $\Omega_{\rm{sat}}$ and the braking constant $K$. Following Bouvier et al.
(\cite{bo97}), $\Omega_{\rm{sat}}$ is fixed to 14\,$\Omega_{\odot}$ and the braking constant $K$
is calibrated on the Sun.
It is beyond the scope of the present paper to compute complete solar models reproducing the detailed solar structure
deduced from helioseismic measurements. In this study, we simply focus on the evolution of the rotation profile
of 1\,M$_{\odot}$ models which reproduce the solar luminosity and radius at the age of the Sun (4.57\,Gyr).

 \begin{figure}[t]
  \resizebox{\hsize}{!}{\includegraphics{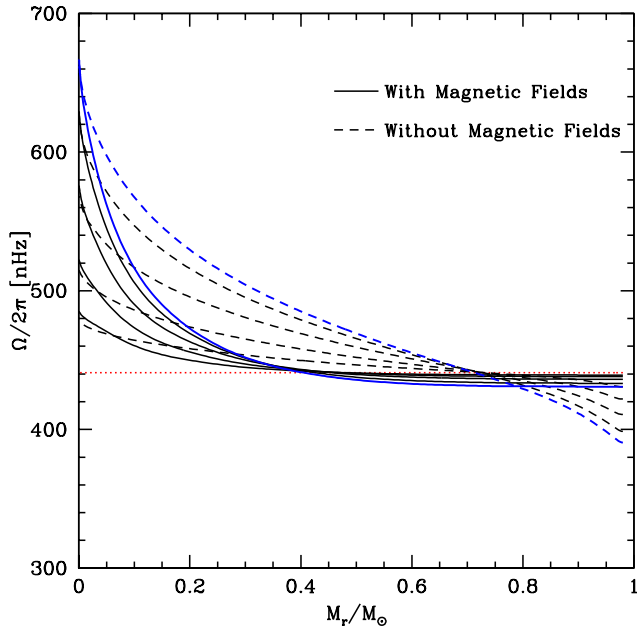}}
  \caption{Rotation profiles
  as a function of the lagrangian mass in solar units for models with (continuous lines) and without magnetic field
  (dashed lines). The dotted line indicates the initial solid body rotation on the ZAMS. 
  The other lines correspond to an age of respectively 1, 2, 3, 4 and 4.57\,Gyr. $\Omega$ increases at the centre during the evolution
  on the main-sequence.
  No magnetic braking at the surface is included.}
\label{pro_no}
\end{figure} 

   \begin{figure}[t]
  \resizebox{\hsize}{!}{\includegraphics{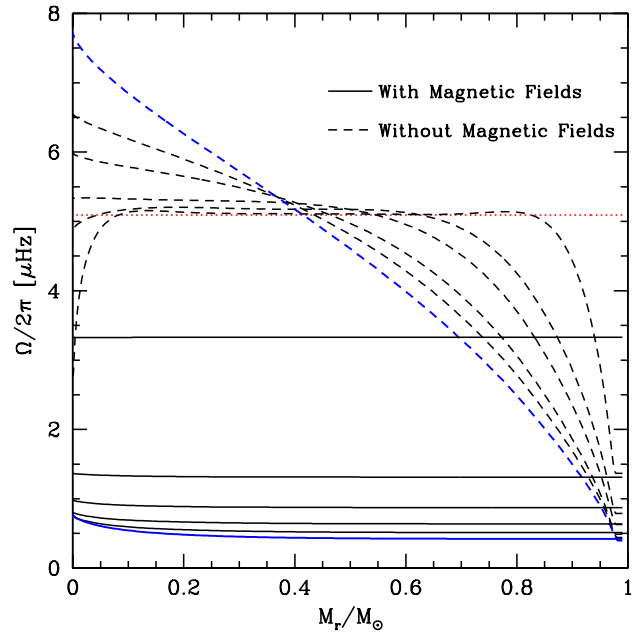}}
  \caption{Same as Fig.~\ref{pro_no} but with an initial velocity of 20\,km\,s$^{-1}$.
           The dotted line indicates the initial solid body rotation on the ZAMS. The other lines correspond to an age of respectively 
	   0.1, 0.5, 1, 2, 3 and 4.57\,Gyr. The surface angular velocity decreases when the star evolves due to the magnetic braking at the surface}
\label{pro_vi20}
\end{figure}

To investigate the effects of magnetic fields on the rotation profile of a slow
rotating solar-like star, we first consider models without magnetic braking at the surface. Two
models are computed: one with rotation only and a second with both rotation and magnetic
field. The initial velocity of these
models is approximately equal to the solar surface rotational velocity.
Fig. \ref{pro_no} shows the evolution of the internal rotation profile
for both models, starting from solid body rotation on the ZAMS. The rotation profile of the model with
rotation only changes with time during the main sequence evolution due to
the transport of angular momentum by circulation. As a result, the model with only rotation
shows after 4.57\,Gyr an angular velocity $\Omega$ which is monotically increasing when the distance to the
centre decreases. The situation is quite different when magnetic fields are accounted for.
As shown in Fig. \ref{pro_no}, the angular velocity $\Omega$ is almost constant 
between the surface and about 0.3\,$M_{\odot}$. 
In the central parts, the angular velocity increases due to the decrease of the horizontal 
coupling insured by the magnetic field
strength resulting from the $\mu$--gradient in the stellar core ($\nu$ varies like $\nabla_{\mu}^{-2}$).

\begin{figure}[t]
  \resizebox{\hsize}{!}{\includegraphics{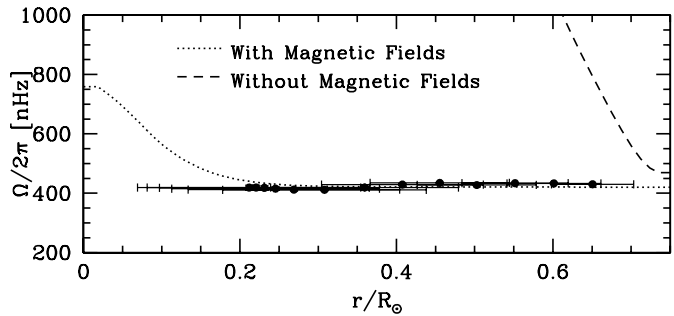}}
  \caption{Rotation profile for a model with rotation only (dashed line) and with both rotation and magnetic field
  (dotted line) at the age of the Sun. The initial velocity is 50\,km\,s$^{-1}$. The points with their respective error bars
  correspond to the angular velocities in the solar radiative zone deduced from GOLF+MDI and LOWL data (Couvidat et al.
  \cite{co03}).}
\label{pro_sol}
\end{figure}

As a second step, models with initial velocities of 20 and 50\,km\,s$^{-1}$ are computed.
The braking law of
Kawaler (\cite{ka88}) is used for these models in order to reproduce the solar surface rotational
velocity. Fig. \ref{pro_vi20} compares the evolution of the internal rotation profile of models with only rotation
and with both rotation and magnetic fields, starting with
$v_{\mathrm{ini}}=20$\,km\,s$^{-1}$ on the ZAMS. We notice that the surface rotation
velocity rapidly decreases due to the magnetic braking at the surface.
For the model including only
rotational effects, this results in a large differential rotation reaching a factor of
about 20 between the angular velocity at the surface and in the stellar core at the age of
the Sun, in good agreement with the previous results of Chaboyer et al. (\cite{ch95}), but in
contradiction with the flat rotation profile of the Sun. 
Fig. \ref{pro_vi20} shows that the model with both rotation and magnetic fields displays
an almost constant angular velocity throughout the radiative interior, with only a small
increase of $\Omega$ in the central parts (for $M_r \leq 0.2$\,$M_{\odot}$). Fig.
\ref{pro_sol}
better compares the theoretical rotation profiles of models computed with an initial
velocity of 50\,km\,s$^{-1}$ to the one deduced from helioseismic measurements (Couvidat et
al. \cite{co03}). Note that the rotation profiles of models with an initial
velocity of 50\,km\,s$^{-1}$ are very similar to those with
$v_{\mathrm{ini}}=20$\,km\,s$^{-1}$. The only difference is that models with
$v_{\mathrm{ini}}=50$\,km\,s$^{-1}$ exhibit slightly faster rotating cores
than those computed with $v_{\mathrm{ini}}=20$\,km\,s$^{-1}$.
Fig. \ref{pro_sol} clearly shows that the rotation profile of models including both
rotation and magnetic fields is in good agreement with the helioseismic measurements,
while models with only rotation
predict a too fast increase of $\Omega$ when the distance to the centre decreases.  

\begin{figure}[tb]
  \resizebox{\hsize}{!}{\includegraphics{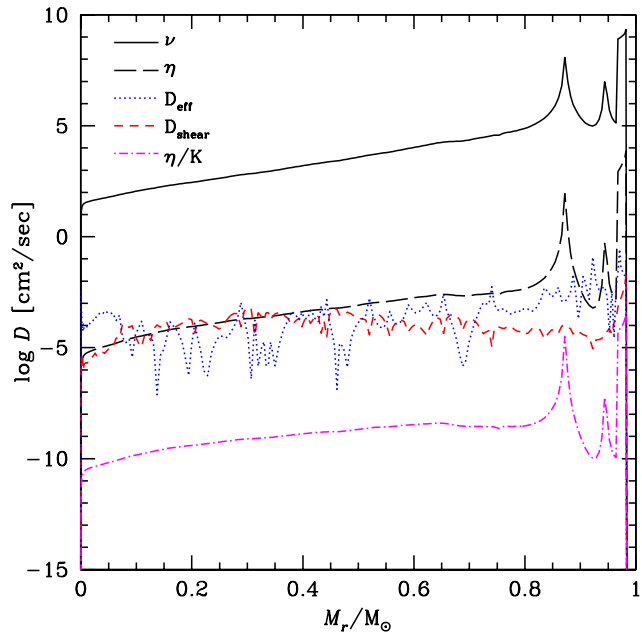}}
  \caption{Diffusion coefficients in the model with both rotation and magnetic fields and $v_{\mathrm{ini}}=50$\,km\,s$^{-1}$ at the age of the Sun.}
\label{cdiff}
\end{figure}

Finally, the values of the various diffusion coefficients for the model with magnetic fields and 
$v_{\mathrm{ini}}=50$\,km\,s$^{-1}$ are shown in Fig. \ref{cdiff}. 
The largest diffusion coefficient
is $\nu$ which acts for the vertical transport of angular momentum. 
The large value of $\nu$
imposes the nearly constant $\Omega$ in the interior and 
confirms the dominant role of the magnetic field for the transport of angular momentum.
The numerical value for the azimuthal component of the field $B_{\varphi}$ is of the order of a few $10^{2}$\,G. 
We also notice the small value of the coefficient of the shear turbulent mixing $D_{\mathrm{shear}}$. 
Indeed, it is of the same order of magnitude as the coefficient $D_{\mathrm{eff}}$, which applies to the transport of chemical elements
by meridional circulation, while in a rotating star without magnetic field $D_{\mathrm{shear}}$ is generally much larger (about
4 orders of magnitude). This small value is a consequence of the near solid body rotation of magnetic models
and suggests that, for slow rotating solar-like stars, the rotation induced mixing is less efficient in magnetic models
than in models with rotation only. This seems to be in agreement with observations.
A detailed study of the evolution of trace elements like lithium and $^{3}$He
with and without magnetic field will be needed to really investigate this point.
Fig. \ref{cdiff} also shows that the value $\eta/K$ is always very small, 
which justifies the simplifications made in deriving Eq. (\ref{equx}).

\section{Conclusion}

The main result of this study is that the Tayler--Spruit dynamo can account
for the flat rotation profile of the Sun as deduced from helioseismic measurements.
There remains however some doubts whether this dynamo is really active in stellar interiors, since 
3D simulations have not yet confirmed the existence and efficiency of this particular instability;
these simulations show a delicate balance between
the generation of the instabilities and their relaxation to stable
configurations (Braithwaite \& Spruit \cite{br04}).
It is also worthwhile to recall that magnetic field is not the only explanation, since 
purely hydrodynamical stellar models including the transport by internal gravity waves constitute
another promising alternative.

\begin{acknowledgements}
We thank S. Couvidat and T. Corbard for providing us with the helioseismic
rotation profile.
We also thank S. Turck-Chi\`eze, C. Charbonnel and S. Talon for usefull discussions.
Part of this work was
supported financially by the Swiss National Science Foundation.
\end{acknowledgements}

\end{document}